\documentstyle[emulateapj,psfig]{article}
\def\w{\omega(\theta)}
\def\A{A_\omega}

\def\eg{{\rm e.g.\ }}

\def\gs{\mathrel{\raise0.35ex\hbox{$\scriptstyle >$}\kern-0.6em 
\lower0.40ex\hbox{{$\scriptstyle \sim$}}}}
\def\ls{\mathrel{\raise0.35ex\hbox{$\scriptstyle <$}\kern-0.6em 
\lower0.40ex\hbox{{$\scriptstyle \sim$}}}}
\def\arcsper{\ifmmode \rlap.{''}\else $\rlap{.}''$\fi}
\def\arcmper{\ifmmode \rlap.{'}\else $\rlap{.}'$\fi}
\slugcomment{Accepted for publication in The Astrophysical Journal (Letters)}

\lefthead{Brainerd \& Smail}
\righthead{A Constant Clustering Amplitude for Faint Galaxies?}

\begin{document}

\title{A Constant Clustering Amplitude for Faint Galaxies?}

\author{Tereasa G.\ Brainerd$^1$ \& Ian Smail$^2$}
\affil{\tiny 1) Boston University, Department of Astronomy, Boston, MA 02215}
\affil{\tiny 2) University of Durham, Department of Physics, Durham, DH1 3LE, UK}

\setcounter{footnote}{2}

\begin{abstract}
The angular clustering of faint field galaxies is investigated using
deep imaging ($I\sim 25$) obtained with the 10-m Keck--I telescope.
The autocorrelation function is consistent with $\w \propto
\theta^{-0.8}$ and, although less steep correlation functions cannot be
ruled out with high confidence, we find no compelling evidence for a
systematic decrease in the power law index at the faintest magnitude
limits.  Results from a number of independent observational studies are
combined in order to investigate the variation of the correlation
amplitude with median $I$-magnitude.  At $I_{\rm med}\sim 23$ the
results obtained by different studies are all in rough agreement and
indicate that for $I_{\rm med} > 22$ the correlation amplitude declines
far less steeply than would be expected from an extrapolation of the
trend in the brighter samples.  In particular, at $I_{\rm med} \sim 24$
our data indicate $\w$ to be a factor $\sim 7$ higher than the
extrapolation.  A near-independence of magnitude is a general feature
of the correlation amplitude in models in which the redshift
distribution of the faint field population contains a substantial
fraction of galaxies with $z\gs 1$.  In order to reproduce the apparent
abrupt flattening of the amplitude of $\w$ observed at faint limits,
approximately 50\% of the galaxies in a sample with a depth of $I\sim
25$ must be at $z > 1$.
\end{abstract}



\section{Introduction}

The angular clustering of faint field galaxies has been studied
extensively (\eg Efstathiou et al.\ 1991, hereafter EBKTG; Roche et
al.\ 1993, 1996; Brainerd, Smail \& Mould 1995, hereafter BSM; Hudon \&
Lilly 1996; Lidman \& Peterson 1996; Villumsen, Freudling \& da Costa
1996; Woods \& Fahlman 1997).  A prime motivation of these studies has
been to investigate the nature of the faint field population.  In
particular, it is possible to infer the effective correlation length of
the sample and the rate at which clustering evolves from a combination
of the amplitude of the angular autocorrelation function, $\w$, and the
redshift distribution of the faint galaxies, $N(z)$.  These
observations can then be used to link properties of the faint field
population with samples of local galaxies. While the exact
interpretation remains controversial, it is generally accepted that
overall $\w$ is fit well by a power law of the form $\theta^{-0.8}$
(although see Infante \& Pritchet (1995) for evidence of a flattening
in the power-law coefficient at faint limits).  Moreover, the amplitude
of $\w$ appears to decline strongly with apparent magnitude, although
it remains significantly non-zero at the faintest limits.  

Some knowledge of the redshift distribution of the galaxies is required
in order to interpret the observed decline of the amplitude of $\w$
with limiting magnitude in terms of the evolution of real-space
clustering.  Although their precise $N(z)$ is not known, the majority
of the faint field population to depths of $I\sim 25$ appear to have
$\left< z \right>\sim 1$ (Kneib et al.\ 1996; Connolly et al.\ 1997).
Given this constraint, the rate of clustering evolution inferred from
$\w$ is consistent with all ``reasonable'' theoretical possibilities.
Similarly, the present-day correlation length of the descendents of the
faint galaxy population is not constrained strongly.  Some authors
argue in favor of a value of $r_0$  close to that of local bright
galaxies (\eg Hudon \& Lilly 1995; Villumsen, Freudling \& da Costa
1996), while others argue for a smaller value, consistent with that
seen for dwarf galaxies (\eg BSM; Infante \& Pritchet 1995).

A feature which has emerged from the studies of $\w$ for faint galaxies
in blue passbands is a slowing of the decline of the correlation
amplitude with apparent magnitude beyond $B\sim 24.5$--25, equivalent
to a median $I$ magnitude of $I_{\rm med}\sim 22.5$ (Roche et
al.\ 1996). A change in the slope of the differential number counts is
also seen in blue passbands at a similar magnitude (Metcalfe et
al.\ 1995; Smail et al.\ 1995).  One interpretation of these features
is that they arise from a steepening in the faint-end slope of the
luminosity function at high redshift, $z\gs 1$ (Metcalfe et
al.\ 1995).  Clearly it is important to confirm and extend these
observations to test this proposal.

Here we present an analysis of $\w$ obtained from deep imaging ($I \sim
25$), sufficient to probe the strength and evolution of galaxy
clustering from $z\sim 1$.  By combining these observations with
results from clustering analyses of other, brighter, $I$-selected
samples we investigate the variation of $\w$ with magnitude.  Using a
modest extrapolation of the observed magnitude-redshift relation of
brighter $I$-selected galaxies we then compare the observed clustering
behavior to theoretical predictions.

\section{Observed Galaxy Clustering}

The data consist of deep $I$-band imaging of two independent fields
(1640+22 and 2229+26) obtained in good conditions with the Low
Resolution Imaging Spectrograph (LRIS, Oke et al.\ 1995) on the 10-m
Keck--I telescope.  The fields are centered on two high galactic
latitude pulsars and thus provide essentially random samples of faint
field galaxies.  Details of the observations, the reduction and
cataloging of the galaxies, and the number counts of the galaxies are
discussed in Smail et al.\ (1995); Reid et al.\ (1996) discuss the
analysis of the faint stars in these fields.  These frames constitute a
superb dataset for the study of the clustering properties of faint
galaxies due to both the excellent seeing ($0.53''$ and $0.58''$ FWHM
for 1640+22 and 2229+26 respectively) and depth (50\% completeness
limits of $I = 26.0$ and $I = 25.6$ for 1640+22 and 2229+26
respectively).   To reduce stellar contamination we consider
only objects with $I \ge 22$ in our analysis.

%
%
\begin{table*}
{\scriptsize
\begin{center}
\centerline{Table 1}
\vspace{0.1cm}
\centerline{Summary of Field Statistics}
\vspace{0.3cm}
\begin{tabular}{lcrcccc}
\hline\hline
\noalign{\smallskip}
{Field} &
{$I$ mag.\ limits} & {$N_{\rm obj}$} & 
{$\delta_{\rm best}$} & {$w(1')^{\,a}$} & $\chi^2(\delta_{\rm best})^{\,b}$ & $\chi^2(\delta=0.8)^{\,b}$\cr
\hline
\noalign{\smallskip}
1640+22........ & 22.0---24.0   &   674 & $0.8$ & $0.038\pm 0.011$ & 0.22 & 0.22 \cr
 & 22.0---24.5 &  989 & $0.6$ & $0.030\pm 0.007$ & 0.30 & 0.39 \cr
 & 22.0---25.0   &  1450 & $0.7$ & $0.028\pm 0.005$ & 0.13 & 0.18 \cr
 & & & & \cr
2229+26........ & 22.0---24.0   &  482  & $1.1$ & $0.050\pm 0.015$ & 0.03 & 0.11 \cr
 & 22.0---24.5 &  699  & $0.9$ & $0.039\pm 0.010$ & 0.26 & 0.31 \cr
 & 22.0---25.0   & 1046  & $0.8$ & $0.035\pm 0.006$ & 0.37 & 0.39 \cr
 & & & & \cr
Combined........ & 22.0---24.0   &  1156  & $0.9$ & $0.043\pm 0.009$ & 0.13 & 0.18\cr
 & 22.0---24.5 &  1688  & $0.7$ & $0.034\pm 0.006$ & 0.31 & 0.32 \cr
 & 22.0---25.0   & 2496  & $0.7$ & $0.031\pm 0.004$ & 0.28 & 0.29 \cr
\noalign{\smallskip}
\noalign{\hrule}
\noalign{\smallskip}
\end{tabular}
\end{center}

\begin{center}
{a)~}{Computed using $\w \propto \theta^{-0.8}$,
corrected for the IC and residual
stellar contamination}
~~~{b)~}{$\chi^2$ per degree of freedom.}
\end{center}
}
\vspace*{-0.8cm}
\end{table*}

The direct pair-counting method proposed by Landy \& Szalay (1993), $\w
= (DD - 2DR + RR)/RR$, was used to estimate $\w$ for all objects with
$22 \le I \le I_{\rm lim}$.  Here $DD$, $DR$, and $RR$ are the number
of unique data-data, data-random, and random-random pairs within a
given angular separation bin centered on $\theta$.  To determine $DR$
and $RR$, mask frames defining areas around bright stars and galaxies
($I < 20$) were constructed for each field and these regions were
excluded from the analysis.  Also excluded was a generous border (15
arcsec) along the frame boundaries, resulting in field sizes of order
30 sq.\ arcmin.  Raw measurements of $\w$ were computed independently
for each field and error bars were assigned to the functions using
bootstrap resampling of the data.  Since the areas of the fields are
almost identical and the residual stellar contamination is similar, it
is fair to compute a mean of the two independent raw measurements of
$\w$ directly.  The result is shown in Fig.~1 where the data points and
error bars are the formal values obtained via a weighted mean of the
corresponding individual functions.

The observed number of objects is used to compute $DD$ and $DR$.  Since
the area of the detector is small, $\w$ is therefore underestimated by
an amount:  ${\rm IC} = \Omega^{-2} \int\int \w d\Omega_1 d\Omega_2$
(\eg Groth \& Peebles 1977), the so-called ``integral constraint''.
Assuming a power law correlation function, $\w = \A \theta^{-\delta}$
with $\delta = 0.8$ as suggested by previous studies, we find ${\rm
IC}_{1640+22} = (0.021\pm 0.002)\A$ and ${\rm IC}_{2229+26} = (0.023\pm
0.002)\A$.

Some stellar contamination remains in our catalogs and results in the
inferred amplitude of $\w$ being lowered from its actual value.
Fortunately the excellent seeing allowed a detailed study of the star
counts in these fields at somewhat brighter magnitudes and from that
analysis a mean stellar contamination can be estimated.  Extrapolating
the star counts obtained by Reid et al.\ (1996) under the assumption
that beyond $I\sim 22$ they remain fairly flat (\eg Fig.~1 of Reid et
al.), we determine a stellar contamination in our catalogs of
$\sim$\,10\% for $I$=22.0---24.0,  $\sim\,$9\% for $I$=22.0---24.5, and
$\sim$\,7\% for $I$=22.0---25.0.  The final corrected amplitude of $\w$
is then given by $\A = \A^{\rm IC} N_{\rm obj}^2 (N_{\rm obj} - N_{\rm
s})^{-2}$, where $N_{\rm obj}$ is the number of objects, $N_{\rm s}$ is
the number of stars, and $\A^{\rm IC}$ is the inferred amplitude of
$\w$ after correcting for the IC.  

The functions in Fig.~1 are consistent with power laws in which $\delta
= 0.8$, as are the individual $\w$ computed for each field.  Formally
the index of the best-fitting power law, $\delta_{\rm best}$, ranges
from 0.6 to 1.1 for the correlation functions computed from the two
fields independently  (Table~1) and from 0.7 to 0.9. for the combined
fields.  There is, however, no clear trend of $\delta_{\rm best}$ with
limiting magnitude, nor is the formal best fit significantly better
than that with $\delta=0.8$.  We therefore adopt $\w = \A\theta^{-0.8}$
for all further analysis.

%
%
\vspace*{-1.3cm}
\hbox{~}
\centerline{\psfig{file=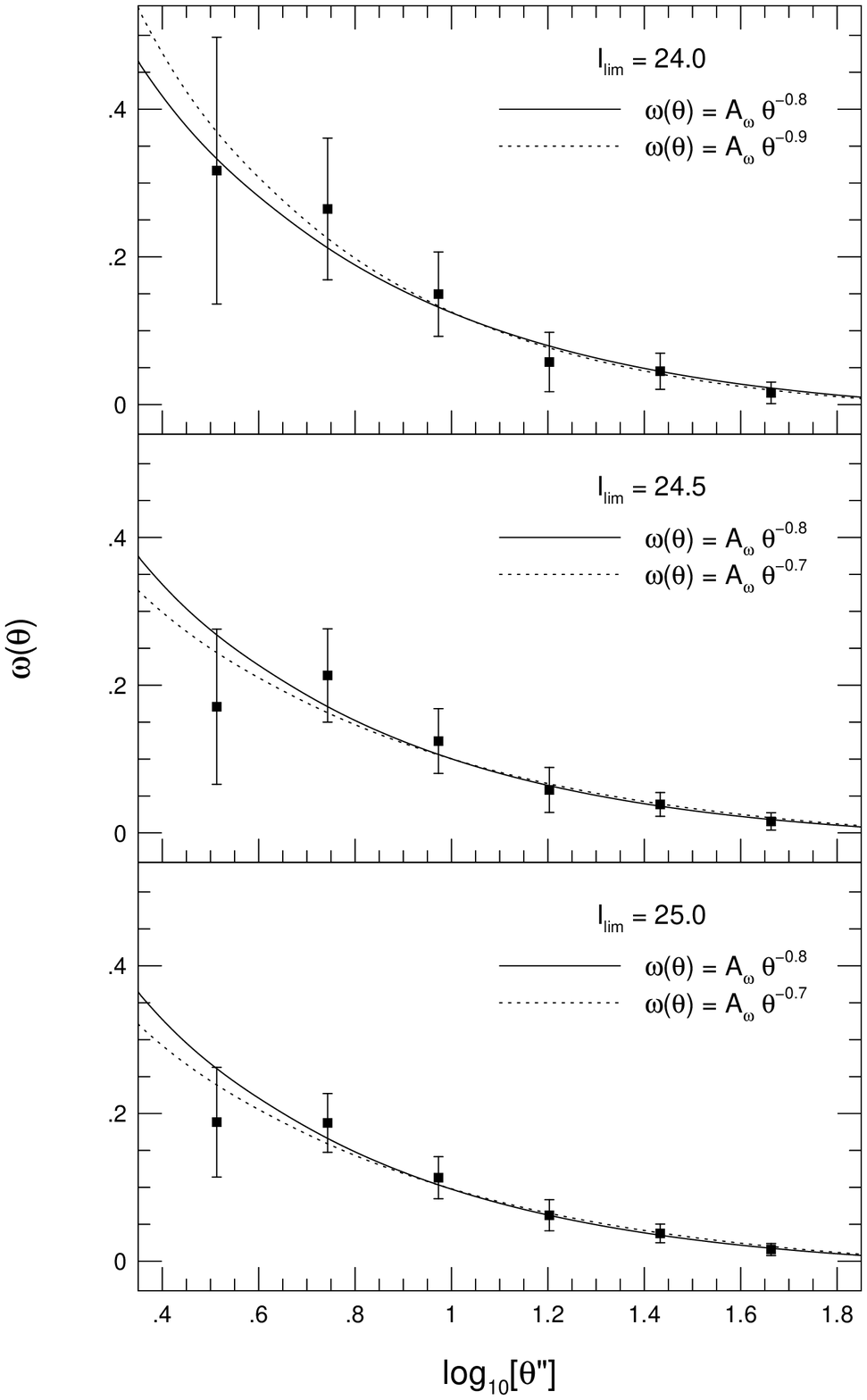,angle=0,width=3.3in}}
\vspace*{-0.7truecm}
\noindent{\scriptsize
\addtolength{\baselineskip}{-3pt} 
Figure~1.
Mean correlation functions obtained by weighted averaging of the raw
functions computed from the two independent fields.  Solid lines
indicate the best-fitting power law of the form $\w = \A\theta^{-0.8}$,
including suppression due to the IC.  Dotted lines show the power
law that is formally the best fit to the observed $\w$. 
See Table~1 for details of the individual fits.   \label{fig1}

\addtolength{\baselineskip}{3pt}
}

Table~1 lists values of $\w$ evaluated at $\theta = 1'$ for the two
individual fields and the combined sample.  The errors listed are the
formal errors derived from the fit of the power law, $\w \propto
\theta^{-0.8}$, to the raw correlation functions. ~From the table,
$\omega(1')$ is nearly independent of limiting magnitude and the
galaxies in 2229+26 are slightly more clustered than the 1640+22
galaxies, although the amplitudes from the two fields agree at better
than 1$\sigma$. Results for $\omega(1')$ as a function of median
$I$-magnitude are plotted in Fig.~2, along with those from other
studies of the clustering of $I$-selected galaxy samples.  

\section{Results and Interpretation}

We start by noting that the clustering amplitudes of the brightest
samples in both our fields are in reasonable agreement with the
clustering of $22 < I < 24$ galaxies reported by EBKTG who find
$\omega(1') = 0.024\pm 0.006$ (corrected for their IC, but not stellar
contamination).  Allowing a mean stellar contamination of order 5\% in
the EBKTG data increases their value to $\omega(1') = 0.027\pm 0.006$
(Fig.~2).  

Performing a weighted linear least squares fit to the data of Lidman \&
Peterson (1996) shown in Fig.~2, we find $\omega(1') \propto I_{\rm
med}^{-0.29}$ at bright magnitudes.  At our faintest limits, however,
$\omega(1')$ is $\sim 7$ times larger than the expectations based on an
extrapolation of the linear fit to the Lidman \& Peterson data.
Moreover, for $I_{\rm med} \gs 22$, $\omega(1')$ appears to be nearly
independent of magnitude, both internally within our sample and in
comparison to slightly brighter studies.  Thus we conclude that the
amplitude of $\w$ for $I$-selected samples of galaxies ceases to
decline steeply with apparent magnitude at $I_{\rm med}\sim 22$ and
remains roughly constant at fainter magnitudes, although a modest rise
or fall in the amplitude at the faintest magnitudes cannot be ruled out.

We construct a simple theoretical model to predict the angular
clustering in our samples and compare this to the observations.  On
small scales ($\theta \ll 1$ rad.) an angular correlation function of
the form $\w = \A \theta^{-\delta}$ corresponds to a spatial
correlation function of the form $\xi(r) = (r/r_0)^{-\gamma}$, where
$\gamma = 1 + \delta$.  Observations of local bright galaxies suggest
that $\xi(r)$ is fit well by a power law with index $\gamma \sim 1.8$
and a correlation length, $r_0$, which varies significantly with both
morphology and luminosity (\eg Loveday et al.\ 1995 and references
therein).  Following Peebles (1980), $\w$ and $\xi(r)$ are related through 
\begin{equation}
\w = \sqrt{\pi} \frac{\Gamma[(\gamma-1)/2]}{\Gamma(\gamma/2)}
     \frac{A}{\theta^{\gamma-1}} r_0^\gamma
\end{equation}
where $A$ is an amplitude factor dependent upon both the shape of
$N(z)$ and the evolution of $\xi(r)$ (see, e.g.\ EBKTG).  The evolution
of $\xi(r)$ can be parameterised by $\xi(r,z) = (r_0/r)^\gamma
(1+z)^{-(3+\epsilon)}$, where $\epsilon = 0.0$ corresponds to
clustering fixed in proper coordinates and $\epsilon = -1.2$
corresponds to clustering fixed in comoving coordinates.  Recent
measurements of $\xi(r,z)$ for $I<22$ galaxies with known redshifts
suggest, however, that the evolution of clustering may have been
moderately rapid and that $\epsilon > 0$.  Le F\`evre et al.\ (1996)
obtain $0 < \epsilon < 2$ for the CFRS galaxies and Shepherd et
al.\ (1997) obtain $\epsilon \sim 1.5$ for a sample of CNOC field
galaxies.

A redshift distribution, $N(z)$, is required in order to make a
theoretical prediction of the clustering amplitude of the faint
galaxies.  As a basis for this we extrapolate from the $N(z)$ observed
for the CFRS galaxies (e.g.\ Le\,F\`evre et al.\ 1996).  For $19 \le I
\le 22$, $N(z)$ for the CFRS galaxies is fit well by $N(z) \propto z^2
\exp[-(z/z_0)^2]$ where $z_0$ is approximately equal to the median
redshift.  Under the assumption that the shape of $N(z)$ does not
change appreciably with depth, median redshifts for our faint galaxies
can be obtained from a linear extrapolation of the CFRS $I_{\rm
med}$--$z$ relation. For galaxies with $22 \le I \le I_{\rm lim}$ we
expect $z_{\rm med} = 0.81, 0.86, 0.91$ for $I_{\rm lim} = 24.0, 24.5,
25.0$ (corresponding to $I_{\rm med} = 23.2, 23.6, 24.0$).  These
predictions are similar to the limits obtained from lensing analyses of
arclets seen through rich clusters of galaxies, which yield $z_{\rm
med} \sim 0.8 \pm 0.1$ for $I<25$ (Kneib et al.\ 1996). 

Using the above shape for $N(z)$ with $z_{\rm med}$ extrapolated from
the CFRS $I_{\rm med}$--$z$ relation, theoretical predictions for
$\omega(1')$ were computed for galaxies with $19 \le I \le I_{\rm
lim}$.  Taking $\gamma = 1.8$, Eq.\ (1) was evaluated at $\theta = 1'$
for $\epsilon = -1.2, 0.0, 1.0$ in two cosmologies: an Einstein-de
Sitter model and an open model with $\Omega_0 = 0.2$.  The shape of the
expected variation of $\log[\omega(1')]$ with $I_{\rm med}$ is
independent of $r_0$ (cf. Eq.\ (1)) and is only weakly dependent upon
the cosmology.  The amplitude of the function, however, is determined
primarily by $r_0$, with a small dependence upon the cosmology. The
theoretical variation of $\omega(1')$ with $I_{\rm med}$ obtained with
$N(z)$ extrapolated from the CFRS results in values of $\omega(1')$ for
a given value of $I_{\rm med}$ in excess of the linear trend suggested
by bright samples, but none of the models reproduces the general trend
of the observational data.  As an illustrative example of this, we show
in ~Fig.\ 2 the predicted variation of $\omega(1')$ for the EdS
universe using $\epsilon = -1.2$ (thin solid line).  In contrast to the
marked flattening suggested by the data, beyond $I_{\rm med} \sim 21$
the predicted amplitude of $\omega(1')$ decreases roughly linearly.
Similar results are obtained for the open model and for $\epsilon =
0.0$ and 1.0.

Due to the reduction of the proper volume element at high redshifts,
the amplitude of $\w$ is expected to become nearly-independent of
magnitude when a high proportion of the galaxies in the sample are at
$z\gs 1$.  Given that the shape of the expected variation of
$\log[\omega(1')]$ with median magnitude is most strongly dependent
upon $N(z)$ and, in particular, upon the tail of the distribution at
large redshift, we modify our fiducial $N(z)$ slightly in order to
attempt to reproduce the general trend of flattening of the observed
correlation amplitude at faint magnitudes.  Given the observational
uncertainties in the determination of the amplitude we do not attempt
to constrain $N(z)$ in a robust manner; rather, we simply address the
question of what fraction of the faint field population must be at
high-$z$ to allow $\omega(1')$ to be nearly-independent of $I_{\rm
med}$ for $I_{\rm med} \gs 23$.  The modification consists of an
extended tail of galaxies at $z=1$--2, which shifts the median
redshifts of our samples to $z_{\rm med} = 0.92, 0.96, 0.98$ for
$I_{\rm lim} = 24.0, 24.5, 25.0$.  At $I_{\rm med} = 24$ this places
$\sim 50$\% of the population at $z > 1$ and $\sim 17$\% at $z > 1.5$,
as opposed to $\sim 44$\% and $\sim 10$\% in the extrapolation of the
CFRS $N(z)$.  

Although slight, this modification of $N(z)$ results in a flattening of
the amplitude of $\w$ at faint limits that is in better agreement with
the trend suggested by the data in Fig.\ 2.  The theoretical curves in
Fig.\ 2 have been normalized to reproduce the observed clustering at
$I_{\rm med} \sim 19$ and for clarity we show only the results for the
EdS universe (but note that the results are similar for the open
universe). In the EdS universe, the normalizations correspond to $r_0 =
4.0, 5.0, 5.9h^{-1}{\rm Mpc}$ for $\epsilon = -1.2, 0.0, 1.0$.  To
obtain identical normalizations in the open model, the correlation
lengths are $r_0 = 4.4, 5.5, 6.5h^{-1} {\rm Mpc}$.  Although none of
the models provide a good fit to all of the data points, the general
flattening trend of the data is reproduced fairly well. 

%
%
\vspace*{-3.truecm}
\hbox{~}
\centerline{\psfig{file=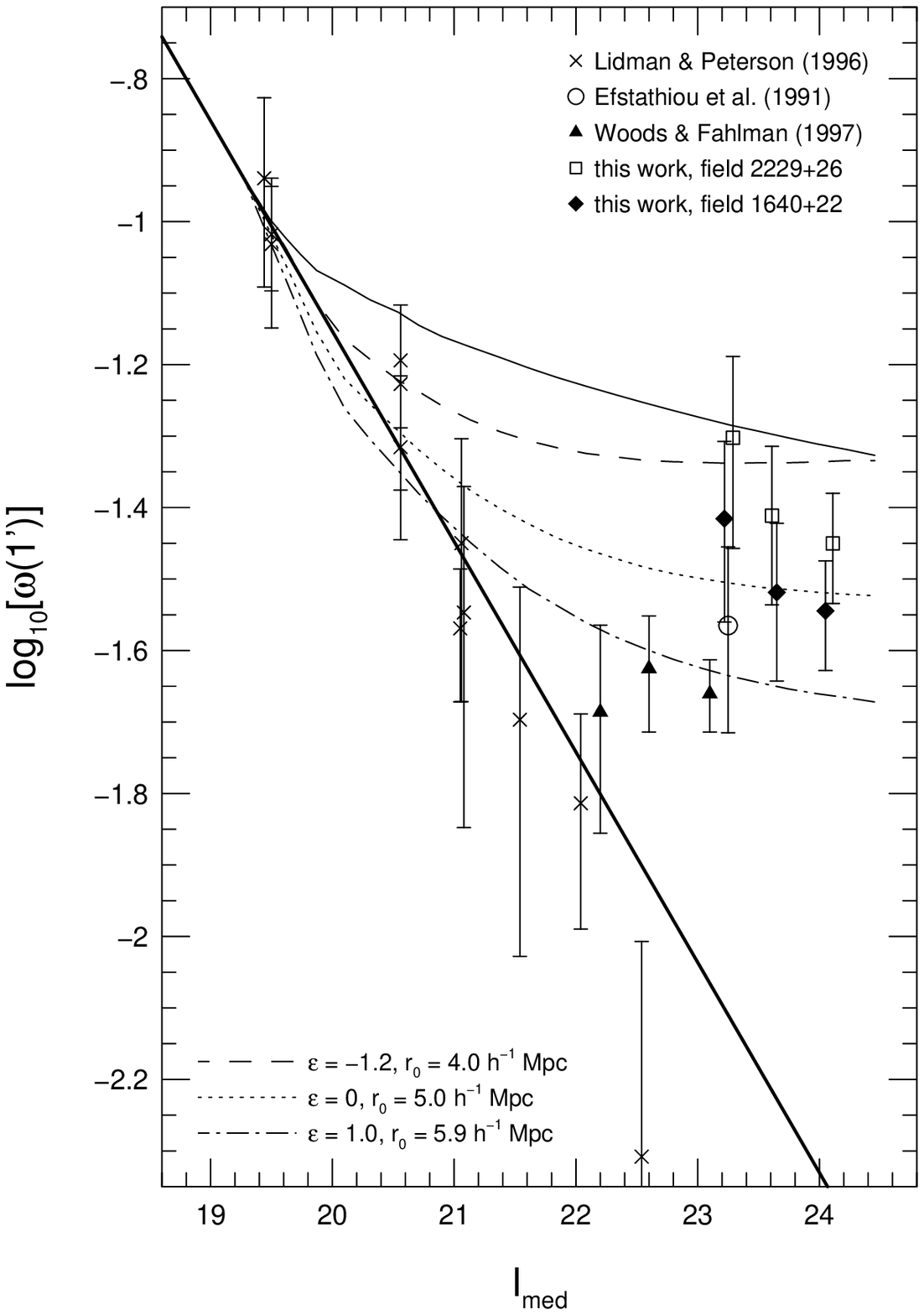,angle=0,width=4.5in}}
\vspace*{-2.truecm}
\noindent{\scriptsize
\addtolength{\baselineskip}{-3pt} 
Figure~2.  Observed variation of $\omega(1')$ with median $I$
magnitude.  All data points have been corrected for the IC and stellar
contamination.  The thick solid line is a weighted linear least squares
fit to the data of Lidman \& Peterson (1996).  Note that systematic
offsets can be introduced into the measured values of $\omega(1')$ by a
number of effects, including atmospheric seeing differences between
datasets, object deblending algorithms, the definition of mask regions,
and stellar contamination corrections.  Such offsets would typically be
comparable in magnitude to the size of the error bars at the faintest
limits.  Also shown is the expected behavior of $\omega(1')$ as a
function of $I_{\rm med}$ for a number of theoretical models in an EdS
universe.  All theoretical predictions have been normalized to
reproduce the observed clustering amplitude at $I_{\rm med} \sim 19$.
The thin solid line shows the theoretical prediction assuming $\epsilon
= -1.2$  and a faint galaxy redshift distribution extrapolated from the
CFRS galaxies (see text).  The dashed, dotted, and dot--dash lines show
the theoretical variation of $\omega(1')$ with $I_{\rm med}$ for
$\epsilon = -1.2$, 0.0, and 1.0 using a faint galaxy redshift
distribution that incorporates a somewhat longer high--$z$ tail than
the simple extrapolation of the CFRS $N(z)$.  \label{fig2}

\addtolength{\baselineskip}{3pt}
}

\section{Discussion and Conclusions}

The interpretation of the observed amplitude of $\w$ in terms of a
correlation length of the galaxy sample is complicated by both a lack
of direct knowledge of the redshift distribution of these galaxies and
the fact that at different limiting magnitudes the sample may be
dominated by different galaxy populations.  It is entirely likely that
the morphological composition of a sample will vary with depth due to
both the shift of the restframe wavelengths sampled by the filter and
the possibility that different galaxy populations may evolve
differently. This may contribute to the different conclusions drawn
from clustering studies of faint galaxies in different passbands.  At
best all that can be concluded based on $\w$ analyses of deep fields is
an effective correlation length and and an effective rate of clustering
evolution since it is clear that the observed clustering is
inextricably linked to the evolution of the galaxy population as a
whole.  Moreover, where fields are specifically chosen to avoid bright
foreground galaxies (\eg BSM) the inferred $r_0$ may be far lower than
the true ``universal'' value since these lines of sight may be biased
toward regions of the sky containing voids and, hence, would have an
uncharacteristically low galaxy clustering amplitude.   Deep wide-field
imaging (sufficient to average over significant amounts of large-scale
structure) is therefore necessary to determine the mean amplitude of
$\w$ accurately at very faint magnitudes and, hence, provide strong
conclusions about the clustering properties of the faint field
population and the reality of any features in the relationship of $\w$
and $I_{\rm med}$.

Here we have measured $\w$ for faint galaxies in two independent random
fields to $I\sim 25$ using high resolution imaging obtained with the
Keck--I telescope.  The angular clustering of the galaxies is
consistent $\w \propto \theta^{-0.8}$, though power laws which are
either somewhat steeper or somewhat shallower cannot be ruled out with
high confidence.  Additionally, the observed clustering amplitude at
our faint limit is consistent with that expected from local bright
galaxies, $r_0\sim 4$--6$h^{-1}$\,Mpc, provided clustering evolves in a
reasonable manner and the redshift distribution extends beyond $z\sim
1$.  This conclusion echos that of Woods \& Fahlman (1997) from their
analysis of an $I_{\rm med} \ls 23$ sample. 

Fainter than $I_{\rm med}\sim 22$ the amplitude of $\w$ is
approximately independent of median magnitude. This result supports
claims of a flattening in the amplitude of $\w$ obtained in blue
samples at a comparable depth (Roche et al.\ 1996).  The apparent
abruptness of the flattening of the amplitude of $\omega(1')$ is not
reproduced well by a model based upon a simple extrapolation of the
$N(z)$ observed for the CFRS galaxies.  We interpret this as evidence
for a modest increase in the fraction of the galaxies at this depth
lying in a high--$z$ tail compared to the extrapolation of the CFRS
$N(z)$. Such a population has also been used to explain the flattening
of the differential number counts in the blue passbands at a depth
equivalent to  $I_{\rm med} \gs 22.5$ (Metcalfe et al.\ 1995; Smail et
al.\ 1995).  The appearance of this feature at the equivalent apparent
magnitude in both blue and red passbands rules out the possibility that
it is caused by a population of very distant galaxies, $z\gs 4$, which
fall out of the sample as the  Lyman-limit moves into the $B$-band.

\section*{Acknowledgments}

We thank Shri Kulkarni and Judy Cohen for their generosity in allowing
us to use the data for this analysis, David Hogg and Lin Yan for help
with the calibration and analysis of the galaxy catalogs, Istv\'an
Szapudi and Henry McCracken for helpful discussions, Doug Sondak for
assistance with code optimization, and the Scientific Computing
Facility at BU for a generous allocation of computing resources.
Support under NSF contract AST-9616968 (TGB) and a PPARC Advanced
Fellowship (IRS) are gratefully acknowledged.  The observations
analysed here were obtained at the W.M.\ Keck Observatory, which is
operated jointly by the California Institute of Technology and the
Univ.\ of California.


\begin{references}
\reference{bsm} Brainerd, T.G., Smail, I. \& Mould, J.R., 1995, \mnras, 275, 781 (BSM)
\reference{con97} Connolly, A., Szalay, A.S., Dickinson, M.,
SubbaRao, M.U. \& Brunner R.J.,  1997, \apjl, 486, L11
\reference{ebktg} Efstathiou, G., Bernstein, G., Katz, N., Tyson, J.A., \&
Guhathakurta, P., 1991, \apjl, 380, L47 (EBKTG)
\reference{gp77} Groth, E.J. \& Peebles, P.J.E., 1977, \apj, 217, 385
\reference{hl96} Hudon, J.D. \& Lilly, S.J., 1996, \apj, 469, 519
\reference{ip95} Infante, L. \& Pritchet, C.J., 1995, \apj, 439, 565
\reference{jpk96} Kneib, J.-P., Ellis, R. S., Smail, I., Couch, W. J., \&
Sharples, R. M.,\apj, 471, 643
\reference{ls93} Landy, S.D. \& Szalay, A.S., 1993, \apj, 412, 64
\reference{ohlcht96} Le F\`evre, Hudon, D., Lilly, S.J., Crampton, D., 
Hammer, F., \& Tresse, L., 1996, \apj, 461,534
\reference{lp96} Lidman, C.E. \& Peterson, B.A., 1996, \mnras, 279, 1357
\reference{lmep95} Loveday, J., Maddox, S.J., Efstathiou, G., \& 
Peterson, B.A., 1995, \apj, 442, 457
\reference{msfr95} Metcalfe, N., Shanks, T., Fong, R., \& 
Roche, N.,  1995, \mnras, 273, 257
\reference{oke95} Oke, J.B., et al.,
1995, \pasp, 107, 375
\reference{jim80} Peebles, P.J.E., 1980, The Large-Scale Structure of the Universe (Princeton:Princeton University Press)
\reference{rymts96} Reid, I.N., Yan, L., Majewski, S., Thompson, I.,  \&
Smail, I., 1996, \aj, 112, 1472
\reference{rsmf93} Roche, N., Shanks, T., Metcalfe, N., \& 
Fong, R., 1993, \mnras, 263, 360
\reference{rsmf93} Roche, N., Shanks, T., Metcalfe, N., \& 
Fong, R., 1996, \mnras, 280, 397
\reference{scye97} Shepherd, C.W., Carlberg, R.G., Yee, H.K.C.,  \&
Ellingson, E., 1997, \apj, 479, 82
\reference{irs95} Smail, I., Hogg, D.W., Yan, L., \& 
Cohen, J.G., 1995, \apjl, 449, L105
\reference{wf97} Woods, D. \& Fahlman, G.G., 1997, \apj, 490, 11
\reference{vfd96} Villumsen, J.V., Freudling, W., \&
da Costa, L.N., 1996, \apj, 481, 578
\end{references}
\end{document}